\documentclass[aps,reprint,prl,superscriptaddress,showpacs,floats,floatfix]{revtex4-2}
\usepackage{graphicx}
\usepackage{amsmath,graphicx,dcolumn}
\usepackage[usenames]{color}
\usepackage{datetime}
\usepackage{textcomp}
\usepackage{ulem}
\usepackage{morefloats}
\usepackage{mathptmx}
\usepackage{indentfirst}
\usepackage{mathrsfs}
\usepackage{bm}
\usepackage{siunitx}
\usepackage{ulem}
\usepackage[colorlinks=true, linkcolor=blue, citecolor=blue, urlcolor=blue]{hyperref}

\newcommand{\blue}{\textcolor{blue}}

\begin{document}
	
\title{Ultrafast Fluence-Reversal Fingerprint of Nonequilibrium Kondo-Hybridized Response in CePt$_2$In$_7$}

\author{Xin-Yi Tian}
\thanks{These authors contributed equally to this work}
\affiliation{School of Physics, Central South University, Changsha 410083, Hunan, China}

\author{Qi-Yi Wu}
\thanks{These authors contributed equally to this work}
\affiliation{School of Physics, Central South University, Changsha 410083, Hunan, China}

\author{Chen Zhang}
\affiliation{School of Physics, Central South University, Changsha 410083, Hunan, China}

\author{Hao Liu}
\affiliation{School of Physics, Central South University, Changsha 410083, Hunan, China}

\author{Yang Luo}
\affiliation{School of Physics, Central South University, Changsha 410083, Hunan, China}

\author{Bo Chen}
\affiliation{School of Physics, Central South University, Changsha 410083, Hunan, China}
	
\author{Ying Zhou}
\affiliation{School of Physics, Central South University, Changsha 410083, Hunan, China}

\author{Zhong-Tuo Fu}
\affiliation{School of Physics, Central South University, Changsha 410083, Hunan, China}

\author{Jin-Dong Bai}
\affiliation{School of Physics, Central South University, Changsha 410083, Hunan, China}

\author{Chun-Hui Lyu}
\affiliation{School of Physics, Central South University, Changsha 410083, Hunan, China}

\author{Zi-Jie Xu}
\affiliation{School of Physics, Central South University, Changsha 410083, Hunan, China}

\author{Hai-Long Deng}
\affiliation{School of Physics, Central South University, Changsha 410083, Hunan, China}

\author{Hai-Yun Liu}
\affiliation{Beijing Academy of Quantum Information Sciences, Beijing 100085, China}

\author{Jun He}
\affiliation{School of Physics, Central South University, Changsha 410083, Hunan, China}
	
\author{Yu-Xia Duan}
\affiliation{School of Physics, Central South University, Changsha 410083, Hunan, China}

\author{Jian-Qiao Meng}
\email{Corresponding author: jqmeng@csu.edu.cn}\affiliation{School of Physics, Central South University, Changsha 410083, Hunan, China}
	
\date{\today}

\begin{abstract}

The nonequilibrium response of a weakly hybridized Kondo lattice remains poorly understood when hybridization competes with magnetic correlations. Here we use femtosecond pump-probe reflectivity to interrogate this problem in the weakly hybridized Kondo-lattice compound CePt$_2$In$_7$. At low fluence, a slow quasiparticle relaxation channel emerges below a crossover temperature $T^* \sim$ 40 K, and its temperature dependence is described phenomenologically by a Rothwarf–Taylor bottleneck with an effective recombination scale 2$\Delta \approx$ 7.4 meV. With decreasing temperature, the coherent-phonon-subtracted nonoscillatory response evolves from an approximately linear fluence dependence to RT-like sublinear behavior near 14–16 K and then to an enhanced high-fluence response at 10 and 6 K. These observations indicate a departure from fixed-gap quasiparticle recombination and provide an ultrafast optical signature of nonequilibrium reorganization in the weakly hybridized Kondo lattice.

\end{abstract}

\maketitle

Heavy-fermion metals pose a fundamental problem in correlated-electron physics: how do nearly localized $f$ moments become part of an itinerant many-body electronic fluid, and how robust is that fluid once it forms? In a Kondo lattice, this transformation is driven by the entanglement between conduction electrons and local $f$ moments, producing heavy quasiparticles, reconstructed Fermi surfaces, and hybridization gaps. The same competition between Kondo screening and Ruderman-Kittel-Kasuya-Yosida exchange also underlies antiferromagnetism, quantum criticality, non-Fermi-liquid behavior, and unconventional superconductivity \cite{PGegenwart2008, SKirchner2020, SDoniach1977}, while the emergence of the heavy-electron fluid follows a universal crossover phenomenology in many Kondo-lattice materials \cite{YFYang2008}. A central unresolved issue is therefore not only where Kondo coherence appears in equilibrium, but whether the resulting low-temperature Kondo-hybridized response behaves as a passive fixed-gap quasiparticle bottleneck or can be dynamically reorganized as the excitation fluence increases.

This distinction is especially important in weakly hybridized Kondo lattices. In such systems, the hybridization gap is not merely an equilibrium energy scale; it reflects a delicate redistribution of $f$-electron spectral weight, magnetic correlations, and quasiparticle entropy, consistent with the broader view that heavy-electron behavior is governed by the progressive emergence and effectiveness of Kondo hybridization \cite{YFYang2012PNAS}. Static probes, including ARPES, scanning tunneling spectroscopy, optical spectroscopy, NMR/Knight-shift measurements, and quantum oscillations, have revealed that $f$-electron itinerancy in heavy-fermion systems is orbital selective, momentum dependent, and probe dependent \cite{PAynajian2012, QYChen2017PRB, QYChen2018PRL, QYChen2019PRL, YZZhao2024SCPMA, YWu2021PRL}; in several Ce- and U-based compounds, it can even partially reverse at the lowest temperatures, with 4$f$/5$f$ spectral weight showing relocalization or back transfer rather than a monotonic buildup of itinerant weight \cite{YLuo2020PRB, PLi2023, FYWu2023, NAzari2023, JJSong2024, NWarren2011PRB}. These observations challenge a simple one-way picture of Kondo coherence as a monotonic crossover into a fixed heavy quasiparticle response. However, equilibrium spectroscopy cannot determine how the low-temperature hybridized electronic structure responds when it is driven out of equilibrium. A decisive test requires a perturbation that acts on the electronic system before lattice thermalization and probes how the associated quasiparticle recombination dynamics recover.

CePt$_2$In$_7$ provides a stringent platform for this test. It is a layered member of the Ce$_m$M$_n$In$_{3m+2n}$ family and orders antiferromagnetically at $T_{\rm N}$ =5.2 K, and becomes superconducting under pressure when magnetism is suppressed \cite{NAWarren2010PRB, HShishido2015JPCS, PHTobash2012JPCM, TKlimczuk2014JPCM, MMansson2014, HSakai2012JPCS, HSakai2011PRB, HSakai2014PRL, EBauer2010}. Compared with the more strongly hybridized Ce-115 compounds, CePt$_2$In$_7$ exhibits weak $c$-$f$ hybridization \cite{YLuo2020PRB, BShen2017CPB} and multiple low-energy scales. NMR suggests a coherence scale near 20 K \cite{HSakai2012JPCS}, whereas Knight-shift analysis points to a higher crossover scale near 40 K and indicates partial 4$f$ relocalization below $\sim$14 K \cite{NWarren2011PRB}. ARPES further reveals an unconventional evolution of Ce 4$f$ spectral weight \cite{YLuo2020PRB}. Previous optical and low-fluence pump-probe measurements established a direct optical feature near 27 meV and an effective quasiparticle relaxation scale of approximately 13 meV \cite{RYChen2016PRB}. These multiple scales make CePt$_2$In$_7$ an ideal platform for testing how its quasiparticle response evolves beyond the recombination regime when weak $c$-$f$ hybridization competes with partial $f$ relocalization and magnetic correlations.

\begin{figure}[tbp]
\vspace*{-0.2cm}
\begin{center}
\includegraphics[width=0.98\columnwidth,angle=0]{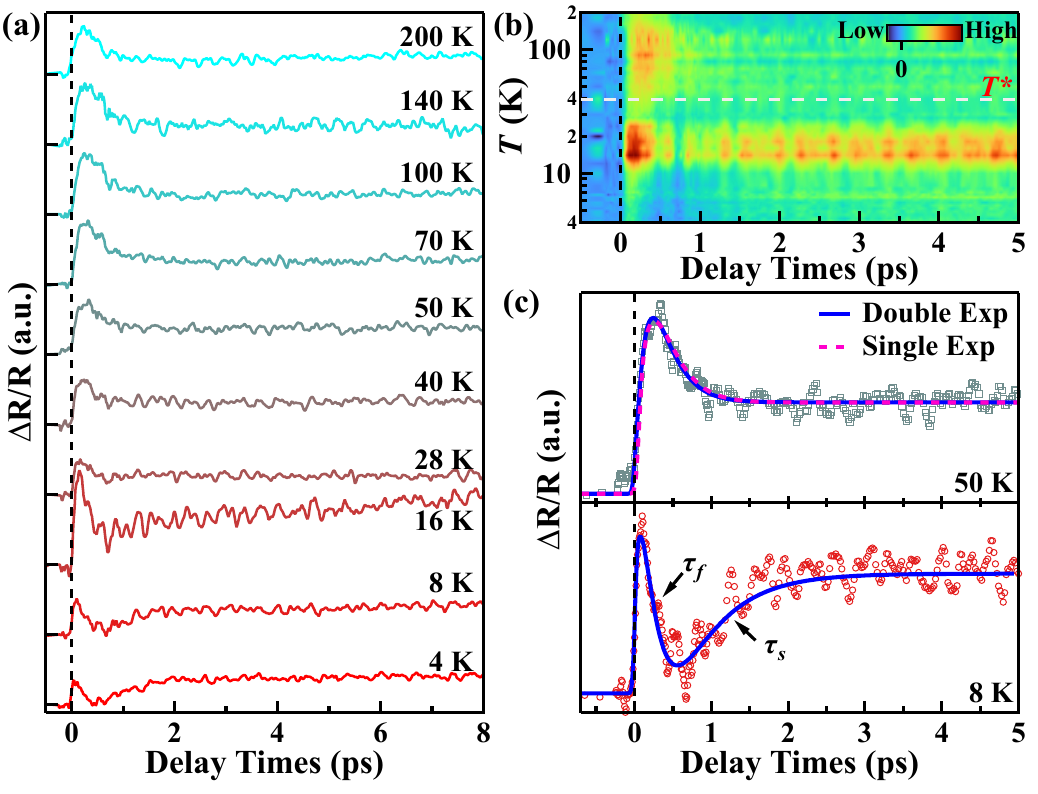}
\end{center}
\vspace*{-0.7cm}
\caption{Emergence of a slow hybridization-related relaxation channel in CePt$_2$In$_7$. (a) Temperature-dependent transient reflectivity $\Delta R/R$ measured at a pump fluence of approximately 15 $\mu$J/cm$^2$. Curves are vertically offset for clarity. (b) Pseudocolor map of $\Delta R/R$ as a function of temperature and delay time. The dashed line marks the crossover scale $T^* \sim$ 40 K, around which the delayed relaxation component gradually becomes resolvable. (c) Representative fits using Eq. (\blue{1}). At 50 K, the slow component is weak and single- and double-exponential fits are nearly indistinguishable. At 8 K, the delayed dip-like response requires the two-component fit. Arrows indicate the fast and slow relaxation time scales, $\tau_f$ and $\tau_s$, respectively.}
\label{FIG:1}
\end{figure}

Ultrafast spectroscopy offers a time-domain route to this issue. In a gapped or pseudogapped quasiparticle spectrum, recombination is constrained by a Rothwarf-Taylor bottleneck: photoexcited quasiparticles and high-frequency bosons exchange energy until the low-energy electronic depletion recovers \cite{JDemsar2003, YPLiu2020PRL, YPei2021PRB,  BLTan2025, YZZhao2023PRB, RYChen2016PRB, JDemsar2006JPCM, DNBasov2005RMP}. The temperature dependence of this channel identifies the relevant recombination scale, whereas its fluence dependence tests whether the same fixed-gap picture remains valid under strong excitation. Such a passive bottleneck should produce sublinear growth, saturation, or gap filling \cite{YZZhao2023PRB, JDemsar2006JPCM}. More generally, strong photoexcitation may smear a gap structure or even drive a photoinduced phase transition \cite{JDemsar2006JPCM}. The distinct signature sought here is a temperature-dependent reversal of the fluence response, rather than a simple monotonic gap filling or gap-smearing effect.

Here we uncover such a nonequilibrium crossover in CePt$_2$In$_7$. Low-fluence dynamics establish a hybridization-related recombination channel below $T^*\sim$ 40 K with a Rothwarf-Taylor scale 2$\Delta \approx$ 7.4 meV. Coherent optical phonons provide an internal lattice reference: their strong equilibrium thermal softening but negligible fluence-induced softening rule out large quasi-equilibrium lattice heating as the dominant origin of the nonlinear response. The central finding is the fluence dependence of $A_{\rm elec}$, the amplitude extracted from the fitted nonoscillatory transient-reflectivity response: with cooling, $A_{\rm elec}$ evolves from nearly linear scaling in the hybridization-crossover regime, to bottleneck-dominated behavior near 14–16 K, and finally to a high-fluence upturn at 10 and 6 K. This fluence reversal exposes a nonequilibrium pathway unavailable to a passive recombination bottleneck, emerging only in the low-temperature relocalization regime of the weakly hybridized Kondo lattice.

High-quality single crystals of CePt$_2$In$_7$ were grown by the In-flux method and freshly cleaved along the (001) surface before optical measurements. Femtosecond pump-probe reflectivity was performed under a vacuum of approximately 10$^{-6}$ mbar using 800-nm ($\sim$1.55 eV) pulses with a duration of about 35 fs from a 1-MHz Yb-based oscillator \cite{XQYe2026, CZhang2022SCPMA, QYWu2025}. The pump and probe spot diameters were approximately 65 and 40 $\mu$m, respectively, with orthogonal polarizations. The transient reflectivity $\Delta R(t)/R$ was measured as a function of delay time, temperature, and pump fluence. Low-fluence temperature-dependent measurements were performed at approximately 15 $\mu$J/cm$^2$, whereas fluence-dependent measurements extended up to 331 $\mu$J/cm$^2$. At the 1.55 eV probe energy, the measured $\Delta R(t)/R$ reflects pump-induced changes in the complex dielectric function, including optical-matrix-element effects. In correlated-electron and Kondo-lattice systems, low-energy electronic reconstruction can influence finite-energy optical responses through self-energy renormalization, changes in carrier scattering, and spectral-weight redistribution linked by the Kramers–Kronig relations \cite{JDemsar2003, JDemsar2006JPCM, DNBasov2005RMP}. A semiquantitative Drude–Lorentz calculation based on the equilibrium optical spectra of CePt$_2$In$_7$ \cite{RYChen2016PRB} further shows that representative changes in low-energy electronic scattering and spectral weight can produce reflectivity variations at 1.55 eV comparable in magnitude to the measured transient response (Supplemental Material Sec. \blue{S8} \cite{SUPPM}).

Figure \blue{1(a)} shows the low-fluence transient reflectivity $\Delta R(t)/R$ from 4 to 200 K. The response contains a prompt electronic component followed by a slower relaxation that becomes increasingly visible upon cooling. The temperature-delay map in Fig. \blue{1(b)} shows that this delayed channel develops continuously below a crossover scale near $T^* \sim$ 40 K. We fit the early-time dynamics using
\begin{equation}
\frac{\Delta R(t)}{R} = (1 - e^{-t/\tau_r})[A_f e^{-t/\tau_f} + A_s e^{-t/\tau_s} + C],
\end{equation}
convoluted with the instrumental response. $\tau_r$ is the rise time, $A_f$ and $\tau_f$ describe the fast component, $A_s$ and $\tau_s$  describe the delayed channel, and $C$ accounts for the long-lived background. At 50 K, single- and double-exponential fits are nearly indistinguishable, whereas at 8 K the delayed dip-like response requires the two-component form [Fig. \blue{1(c)}]. The full temperature-dependent data are shown in Supplemental Material Sec. \blue{S1} \cite{SUPPM}. The fast component remains shorter than approximately 0.3 ps and is naturally assigned to the initial redistribution of photoexcited carriers through electron-boson scattering, as discussed in Supplemental Material Sec. \blue{S2} \cite{SUPPM}. We therefore focus on the slow component as an optical signature of low-energy quasiparticle recombination associated with weak $c$-$f$ hybridization.

The slow channel carries the dynamical fingerprint of weak $c$-$f$ hybridization and its interplay with magnetism. Figure \blue{2} shows that  both $|A_s|$ and $\tau_s$ become resolvable below $\sim$ 40 K and develop a pronounced nonmonotonic structure: $|A_s|$ grows upon cooling, reaches a maximum near $\sim$ 14 K, is suppressed on approaching $T_{\rm N}$, and then increases again at the lowest temperatures. The onset near $T^* \sim$ 40 K is consistent with the $^{\rm 195}$Pt Knight-shift anomaly associated with the emergence of a Kondo-liquid component in CePt$_2$In$_7$ \cite{NWarren2011PRB}, while the change below $\sim$ 14 K coincides with the partial relocalization of hybridized quasiparticles reported by the same Knight-shift measurements \cite{NWarren2011PRB}. Transport measurements also show changes in a nearby low-temperature range, including a resistivity shoulder around $\sim$ 25 K and a Hall-coefficient maximum near $\sim$ 20 K, indicating that the low-temperature electronic scattering and transport properties evolve well above $T_{\rm N}$ \cite{PHTobash2012JPCM}. The overall magnitude and broad temperature evolution of $\tau_s$ are consistent with the previously reported slow relaxation channel \cite{RYChen2016PRB}, while denser low-temperature sampling resolves additional structure in both $|A_s|$ and $\tau_s$. Related ultrafast studies show that magnetic order can affect relaxation amplitudes and lifetimes: $A_2$ and $\tau_2$ show anomalies at $T_{\rm N}$ in CeAuSb$_2$ \cite{YZZhao2023PRB}, and time-resolved photoemission in UPd$_2$Al$_3$ reveals relaxation dynamics modified by the antiferromagnetic state \cite{JJSong2026}. Thus, the slow channel in CePt$_2$In$_7$ reflects recombination in a weakly hybridized Kondo lattice whose dynamics are influenced by partial 4$f$ relocalization and magnetic correlations.

\begin{figure}[tbp]
\vspace*{-0.2cm}
\begin{center}
\includegraphics[width=0.98\columnwidth,angle=0]{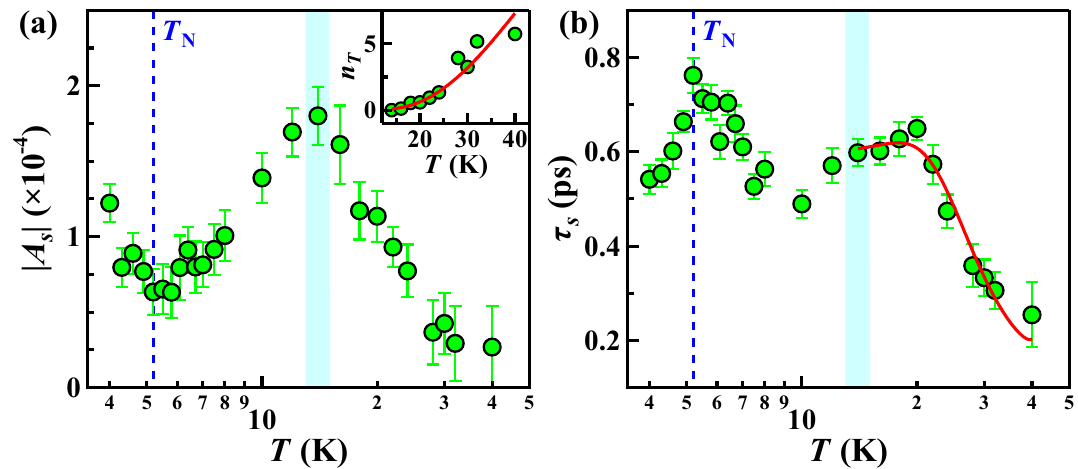}
\end{center}
\vspace*{-0.7cm}
\caption{Rothwarf-Taylor bottleneck analysis of the slow relaxation channel. (a) Temperature dependence of the slow-component amplitude $|A_s|$. The inset shows the thermally excited quasiparticle density $n_T$ extracted from $|A_s|$. (b) Temperature dependence of the slow relaxation time $\tau_s$. Red solid curves are Rothwarf-Taylor fits, yielding a low-energy bottleneck scale 2$\Delta \approx$ 7.4 meV. The shaded region indicates the low-temperature reconstruction scale near $\sim$14 K.}
\label{FIG:2}
\end{figure}

To quantify the recombination scale, we analyze the slow component using a Rothwarf-Taylor bottleneck framework \cite{JDemsar2006JPCM, ARothwarf1967, Kabanov1999}. The relaxation rate and thermally excited quasiparticle density are written as
\begin{equation}
\tau^{-1}(T) = \Gamma [\delta (\epsilon n_T + 1)^{-1} + 2 n_T],
\end{equation}
\begin{equation}
n_T(T) = \frac{A(0)}{A(T)}-1\propto(T\Delta)^{p}e^{(-\Delta/T)},
\end{equation}
where $n_T$ is the thermally excited quasiparticle density, $\Gamma$, $\delta$, and $\epsilon$ are temperature-independent parameters, and $p$  describes the density-of-states profile near the depletion edge. Taking $p$ = 0.5, appropriate for a BCS-like density-of-states profile \cite{JDemsar2006JPCM}, yields 2$\Delta \approx$ 7.4 meV within the specified fitting interval. The difference from the previously reported pump-probe estimate  \cite{RYChen2016PRB} may reflect the different fitting intervals, low-temperature sampling, and the additional constraint provided here by the amplitude-derived $n_T(T)$. The extracted 2$\Delta$ should be interpreted as a low-energy recombination scale, not as the full direct optical hybridization gap \cite{RYChen2016PRB}. The distinction is essential: time-resolved reflectivity is sensitive to the near-$E_F$ quasiparticle depletion that controls recombination, whereas infrared spectroscopy probes the direct optical transitions across the hybridization gap. The crossover near $T^* \sim$ 40 K agrees with the Knight-shift scale and lies above the NMR coherence scale \cite{HSakai2012JPCS, NWarren2011PRB}, showing that different probes access different aspects of the gradual Kondo-coherence crossover in CePt$_2$In$_7$.

\begin{figure}[tbp]
\vspace*{-0.2cm}
\begin{center}
\includegraphics[width=0.90\columnwidth,angle=0]{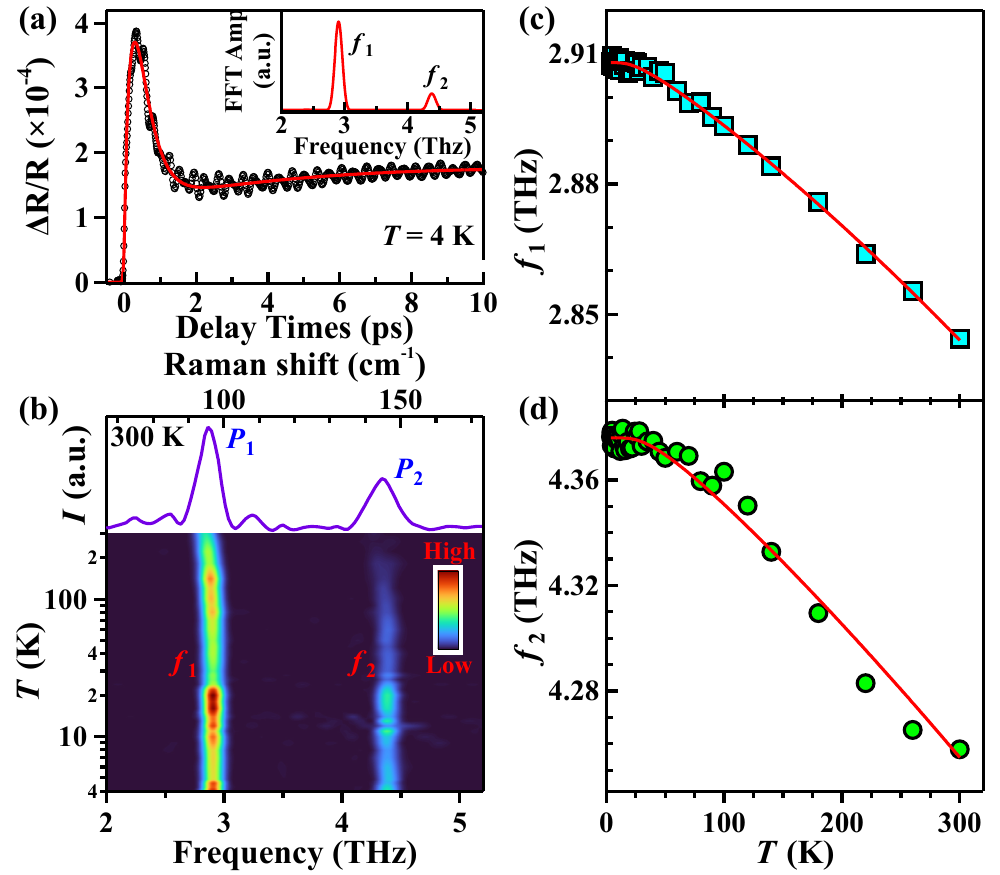}
\end{center}
\vspace*{-0.7cm}
\caption{Coherent optical phonons as an internal lattice reference. (a) Representative high-fluence transient reflectivity at 4 K. The red curve is the fitted nonoscillatory electronic background. The inset shows the Fourier spectrum of the residual oscillatory component, with coherent phonon modes at $f_1 \approx$ 2.91 THz and $f_2 \approx$ 4.38 THz. (b) Raman spectrum measured at 300 K and temperature-dependent FFT intensity map of the time-domain oscillations. The Raman modes near 95.5 and 144.7 cm$^{-1}$ match the coherent phonons observed in pump-probe reflectivity. (c,d) Temperature dependence of $f_1$ and $f_2$, respectively. Red curves are fits using an anharmonic phonon model.}
\label{FIG:3}
\end{figure}

\begin{figure*}[tbp]
\vspace*{-0.2cm}
\begin{center}
\includegraphics[width=1.6\columnwidth,angle=0]{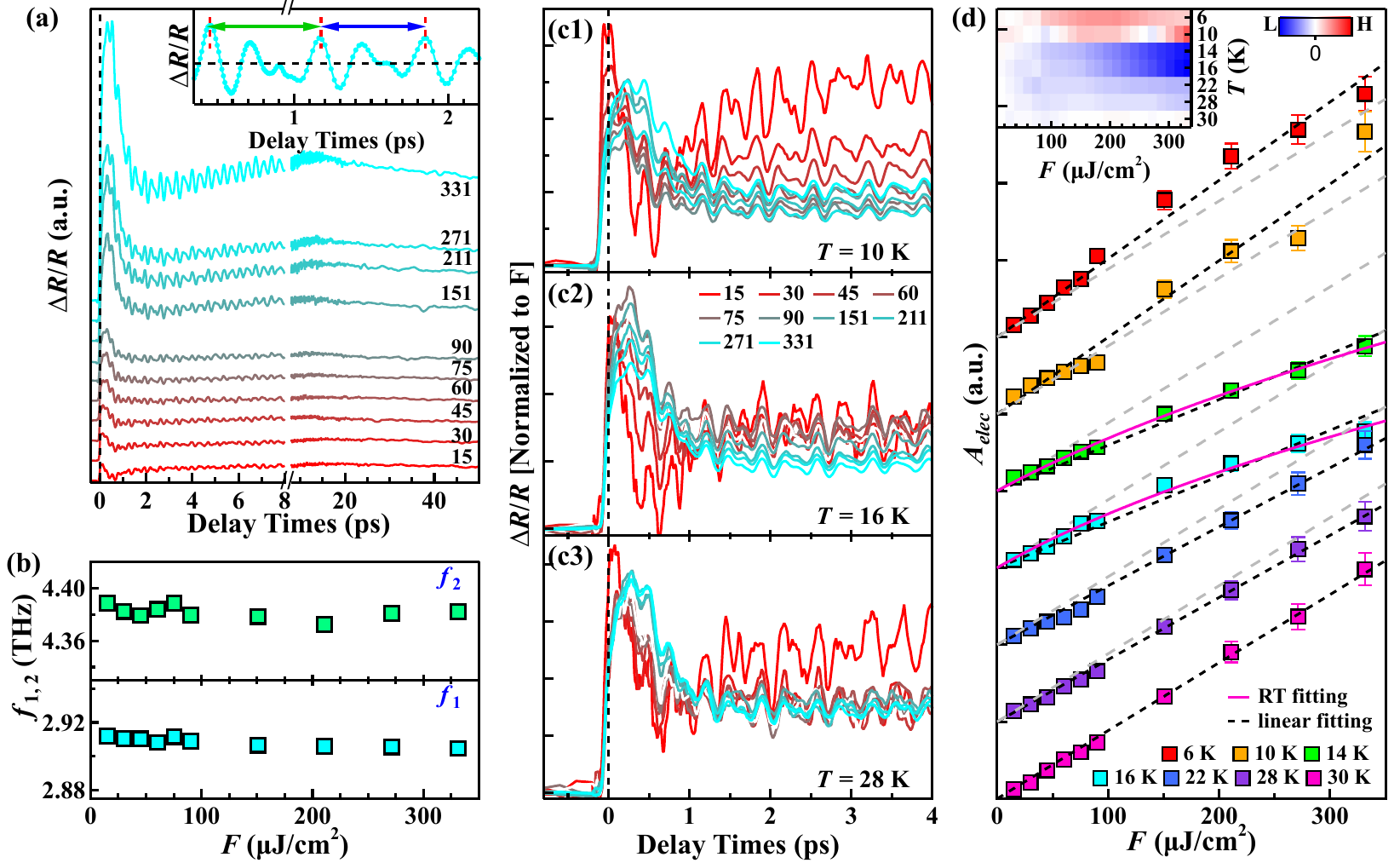}
\end{center}
\vspace*{-0.7cm}
\caption{Fluence-dependent reversal of the Kondo-hybridized electronic response. (a) Fluence-dependent transient reflectivity $\Delta R/R$ at 10 K. Curves are vertically offset for clarity. The inset shows the residual oscillatory component at 331 $\mu$J/cm$^2$ after subtracting the nonoscillatory electronic background. (b) Fluence dependence of the coherent phonon frequencies $f_1$ and $f_2$, showing negligible fluence-induced softening. (c1-c3) Fluence-normalized transient reflectivity ($\Delta R/R$)/$F$ at 10, 16, and 28 K. The lack of collapse demonstrates intrinsic nonlinear fluence dependence. (d) Fluence dependence of the fitted nonoscillatory amplitude $A_{\rm elec}$ at 6, 10, 14, 16, 22, 28, and 30 K. Black dashed lines are linear fits; gray dashed lines are vertically shifted copies of the 30 K linear response; magenta solid curves are Rothwarf-Taylor fits to the 14 and 16 K data using Eq. (\blue{4}). The inset shows the deviation map $\delta A(T, F) = A_{\rm elec}(T, F) - A_{\rm elec}^{linear}(30, F)$. The response evolves from a weak-linear regime at 30, 28, and 22 K, through a representative bottleneck-dominated sublinear regime at 14 and 16 K, to a low-temperature high-fluence enhancement regime at 10 and 6 K. Data sets at different temperatures are vertically offset for visual clarity.}
\label{FIG:4}
\end{figure*}

A central issue in any ultrafast experiment on heavy fermions is whether the observed response is electronic or merely thermal. Because the quasiparticle effective mass is large and the low-energy scales are small, even modest optical excitation can transiently elevate the electronic temperature. We therefore separate three stages of the nonequilibrium process. First, the pump pulse creates a nonthermal carrier distribution that rapidly redistributes energy within the electronic and strongly coupled bosonic subsystems. Second, recombination across the low-energy hybridization-related depletion proceeds through a bottleneck, captured by the slow channel. Third, the lattice approaches a quasi-equilibrium temperature on longer time scales. In the language of a two- or three-temperature model, the electronic system and a subset of strongly coupled bosons can be driven far from equilibrium before the full lattice bath is appreciably heated; in heavy-fermion systems, this separation is further amplified by the anomalous electron-phonon relaxation associated with the heavy quasiparticle density of states \cite{KHAhn2004}.

Coherent phonons provide an internal lattice clock for tracking this separation. Figure \blue{3(a)} shows a representative high-fluence transient trace at 4 K. After subtracting the nonoscillatory electronic background, the residual oscillatory component contains two coherent modes at $f_1 \approx$ 2.91 THz and $f_2 \approx$ 4.38 THz. Raman spectroscopy at 300 K resolves modes near 95.5 and 144.7 cm$^{-1}$ (see details in Supplemental Material Sec. \blue{S3} \cite{SUPPM}), corresponding to approximately 2.86 and 4.34 THz [Fig. \blue{3(b)}], in good agreement with the time-domain modes. These oscillations are therefore coherent optical phonons rather than collective amplitude modes of an electronic order parameter.

The phonon frequencies are strongly temperature dependent: $f_1$ softens from approximately 2.91 to 2.84 THz and $f_2$ from approximately 4.38 to 4.26 THz between 4 and 300 K [Figs. \blue{3(c)} and \blue{3(d)}]. Together with the damping rates shown in Supplemental Material Sec. \blue{S4} \cite{SUPPM}, this evolution is well described by a standard anharmonic phonon model \cite{Balkanski1983, Menendez1984}. By contrast, both phonon frequencies remain nearly unchanged with increasing pump fluence up to 331 $\mu$J/cm$^2$ [Fig. \blue{4(b)}]. Although their weak thermal dispersion at the lowest temperatures limits sensitivity to modest temperature changes, this fluence independence, together with the distinct temperature-dependent electronic response, argues against substantial quasi-equilibrium lattice overheating. The coherent phonons therefore provide an internal reference for the lattice response while allowing for transient nonequilibrium within the electronic and strongly coupled bosonic subsystems.

We now turn to the fluence dependence, which is the key nonequilibrium test. Figure \blue{4(a)} shows the transient reflectivity at 10 K. At low fluence, the signal retains the delayed dip-like recovery characteristic of the low-temperature hybridized state. With increasing fluence, the early dip is suppressed and a stronger delayed electronic component develops. The residual oscillatory signal also shows a delay-dependent period at high fluence, indicating that the photoexcited electronic potential transiently renormalizes the coherent lattice motion. This is precisely the regime in which electronic, bosonic, and lattice degrees of freedom are not yet described by a single temperature.

To isolate the electronic part of the response, we define $A_{\rm elec}$ as the peak amplitude of the fitted nonoscillatory background after removing the coherent phonon contribution. This definition minimizes contamination from the lattice oscillations and is tested by the background-subtracted quantity ($A_{\rm elec}-C$) in Supplemental Material Sec. \blue{S7} \cite{SUPPM}. The fluence-normalized traces in Figs. \blue{4(c1)}-\blue{4(c3)} do not collapse, demonstrating that the transient reflectivity is intrinsically nonlinear in fluence.

The physical content of this nonlinearity is summarized by $A_{\rm elec}(F)$ in Fig. \blue{4(d)}. The key new observation is the fluence-dependent reversal of this phonon-free electronic response, which separates three nonequilibrium regimes. At 30 K, the response is nearly linear in fluence and provides the primary weak-hybridization reference. The 28 and 22 K data are also well described by linear functions, although they lie slightly below the 30 K reference. Thus, cooling within this regime does not yet produce a fully nonlinear bottleneck; instead, it gradually reduces the effective electronic susceptibility, consistent with the growth of a weak hybridization-related depletion of low-energy states.

Upon further cooling, the response enters a bottleneck-dominated regime. The 14 and 16 K data show representative sublinear fluence dependences and are well described by the Rothwarf-Taylor form \cite{JDemsar2006JPCM}
\begin{equation}
A \propto \sqrt{1 + kF} - 1,
\end{equation}
where $k$ is a constant. We use these temperatures as representative examples of the intermediate-temperature bottleneck regime, not as the only temperatures that may follow an RT-like dependence. The agreement with Eq. (\blue{4}) indicates that additional pump fluence produces fewer effective low-energy quasiparticles than expected for a linear metallic response, because recombination is constrained by a depleted hybridization-related density of states. In nonequilibrium thermodynamic terms, the electronic subsystem does not simply absorb energy as a featureless metal; it stores and releases energy through a restricted quasiparticle-boson recombination channel.

At the lowest temperatures, the fluence dependence changes qualitatively. At 10 and 6 K, the high-fluence data rise above both the weak-linear reference and the RT-like sublinear trend. A passive fixed-gap bottleneck would instead give sublinear growth, saturation, gap filling, or gap smearing under strong excitation \cite{YZZhao2023PRB, JDemsar2006JPCM}. The upturn is also inconsistent with simple lattice heating, which would drive the low-temperature trace toward the higher-temperature weak-hybridization response. Below about 14 K, Knight-shift measurements indicate partial relocalization of the hybridized quasiparticles as a precursor to their participation in antiferromagnetic order \cite{NWarren2011PRB}. We therefore associate the high-fluence enhancement with a nonequilibrium electronic channel active in this relocalization regime, where strong excitation perturbs the balance among Kondo screening, partial 4$f$ relocalization, and magnetic correlations.

This interpretation is supported by three consistency checks. First, the fluence reversal survives subtraction of the long-lived background $C$, as shown by ($A_{\rm elec}-C$) and $\delta(A_{\rm elec}-C)$ in Supplemental Material Sec. \blue{S7} \cite{SUPPM}. Second, using either the 30 K or 28 K linear response as the weak-linear reference yields the same qualitative evolution: 22 and 28 K remain approximately linear, 14 and 16 K show bottleneck-like sublinear suppression, and 10 and 6 K show enhanced high-fluence responses. Independent RT-plus-linear and RT-plus-quadratic analyses in Supplemental Material Sec. \blue{S7} \cite{SUPPM} further show that the additional contribution beyond the RT bottleneck is minimized at 14–16 K and reemerges at 10 and 6 K. Third, the coherent-phonon frequencies remain nearly independent of fluence, arguing against substantial quasi-equilibrium lattice overheating. These checks demonstrate that the observed evolution is robust against background subtraction, reference-temperature choice, and fitting parametrization.

The resulting physical picture is the following. Above the relocalization scale, photoexcitation probes a weakly hybridized but nearly linear electronic response. Near the maximum of the slow-channel amplitude, recombination is bottlenecked by a low-energy hybridization-related depletion. At lower temperatures, where partial 4$f$ relocalization and magnetic correlations become increasingly relevant, strong excitation activates an additional pathway beyond simple filling or saturation. Time-resolved reflectivity does not directly measure the momentum-resolved 4$f$ spectral function; nevertheless, the robust fluence dependence of $A_{\rm elec}$ provides an optical signature consistent with a change in the nonequilibrium response of the low-temperature Kondo-hybridized state.

In summary, femtosecond pump-probe reflectivity shows that weak Kondo hybridization in CePt$_2$In$_7$ cannot be captured by a passive fixed-gap recombination picture. Low-fluence dynamics identify a hybridization crossover near $T^*\sim$ 40 K and a Rothwarf-Taylor recombination scale 2$\Delta \approx$ 7.4 meV, while coherent phonons rule out large quasi-equilibrium lattice heating as the origin of the nonlinear fluence dependence. The key observation is that $A_{\rm elec}(F)$ evolves from nearly linear scaling, through bottleneck-dominated suppression near 14–16 K, to high-fluence enhancement at 10 and 6 K. This evolution establishes fluence-dependent pump-probe spectroscopy as a sensitive probe of how the weak heavy-fermion hybridization reorganizes in the partial-relocalization regime.

This work was supported by the National Natural Science Foundation of China (Grants No. 12574168 and No. 12074436), the National Key Research and Development Program of China (Grant No. 2022YFA1604204), the Beijing National Laboratory for Condensed Matter Physics (Grant No. 2024BNLCMPKF001), and the Science and Technology Innovation Program of Hunan Province (Grant No. 2022RC3068).

\end{document}